\documentclass[aps,amsmath,amssymb,superscriptaddress,prb,showpacs]{revtex4}
\usepackage[dvips]{graphicx}
\usepackage{psfrag}

\newcommand{\be}{\begin{equation}}
\newcommand{\e}{\end{equation}}
\newcommand{\beml}{\begin{subequations}}
\newcommand{\eml}{\end{subequations}}
\newcommand{\beq}{\begin{eqnarray}}
\newcommand{\eq}{\end{eqnarray}}
\newcommand{\ba}{\begin{array}}
\newcommand{\ea}{\end{array}}

\begin{document}

\author{S. Flach}
\affiliation{Max-Planck-Institut f\"ur Physik komplexer Systeme, N\"othnitzer
Strasse 38, D-01187 Dresden, Germany}

\author{V. Fleurov}
\affiliation{Raymond and Beverely Sackler Faculty of Exact Sciences,
School of Physics and Astronomy,
Tel Aviv University,
Tel Aviv 69978, Israel}

\author{A.V. Gorbach}
\affiliation{Max-Planck-Institut f\"ur Physik komplexer Systeme, N\"othnitzer
Strasse 38, D-01187 Dresden, Germany}

\title{Classical and quantum radiation of perturbed discrete breathers}
\date{\today}

\begin{abstract}
We show that the linearized phase space flow around a discrete
breather solution is not capable of generating persistent
energy flow away from the breather even in the case of instabilities
of extended states. This holds both for the classical and quantized
description of the flow. The main reason for that is the
parametric driving the breather provides to the flow.
Corresponding scaling arguments are derived for both
classical and quantum cases. Numerical simulations of the classical
flow support our findings.
\end{abstract}


\maketitle

\section{Introduction}
Discrete breathers (DB) have been intensively studied in the
past decade. They are known to be generic solutions
of the dynamics of nonlinear spatially discrete translationally invariant
Hamiltonian
systems. DBs are time-periodic and spatially
localized excitations and belong to one-parameter families
of solutions of the underlying equations of motion \cite{DB_rev}.

DBs have been observed in various experimental situations
ranging from Josephson junction ladders \cite{JJL}, coupled nonlinear
optical waveguides \cite{OW} and driven micromechanical cantilever arrays
\cite{MCA}
to layered antiferromagnets \cite{magnet} and
high-$T_c$
superconductors \cite{hightc},
surface and bulk lattice vibrations
of solids \cite{lattice} and Bose-Einstein
condensates loaded on optical lattices \cite{BEC}.
DBs are also predicted
to exist in the dynamics of dusty plasma crystals \cite{plasma}.
The characteristic spatial scales
range from micrometers down to Angstroems.
Especially in Josephson junction ladders DB excitations have been
studied very extensively, including their interaction with
the modes of the lattice part which is not excited by the DB
\cite{JJL_Fano}.
Thus the issue of stability of DB states, of their interaction with
lattice modes, and possible mechanisms of radiation of energy
by DBs due to this interaction becomes an important and timely
issue.

The fact that DB solutions are generic for nonlinear Hamiltonian lattices
implies that in the majority of cases the underlying Hamiltonian
equations of motion are not integrable. This puts limitations
and complications on the study of perturbed DB states.
If the perturbation is considered to be of small amplitude,
its evolution can be described using a linearization of the phase space
flow around the DB. The resulting linear coupled ordinary differential
equations (ODE) with time-periodic coefficients can be studied within the
framework of Floquet theory
\cite{Flach_Book}.

Taking into account higher order terms
in the phase space flow will ultimately lead to more complicated nonintegrable
equations, which can be studied only approximatively
\cite{Johan_Aubry}.
Several issues are at stake when discussing the evolution of perturbed
breathers. One can consider localized or extended
perturbations on one hand. On the other hand there
are differences in the way breathers react to such perturbations
depending on the amplitude of the latter.

Let us first discuss the case of a linearized phase space flow
around a DB. Formally the obtained Floquet equations decouple
the dynamics of the DB (which is assumed to be given) from 
the evolution of the perturbation. Perturbations which grow in time
will then invalidate the abovementioned linearization.
Perturbations which decay in time do not contradict the linearization,
but in fact the energy stored in the initial perturbation cannot
simply disappear if we consider Hamiltonian dynamics. Consequently
there is a subtle way this energy will have to be transferred
to the DB, which again is beyond the linearization frame.
Marginally stable perturbations, which neither grow nor decay seemingly do not
violate the assumed linearization. Nevertheless it has been shown
that for extended perturbations such a case may be accompanied
by a nonzero energy flux emitted out of or into the breather core
\cite{Cretegny}.
Again it would violate the linearization frame.
Even though it may do so, such predicted radiation scenaria
are confirmed in numerical simulations, underpinning the use
of the linearization picture.

Additional sources of radiation can appear when
taking into account nonlinear corrections to the phase space flow
of a DB perturbation,
even if the linearized case did not provide with such sources.
For instance a marginally stable localized linearized perturbation
will yield an energy radiation due to the appearance of new frequency
combinations and resonances with the spectrum of small amplitude
plane waves \cite{DB_decay}.

Here we will be concerned with a particular case within
the linearized phase space flow frame, which corresponds
to the abovementioned extended perturbations yielding
a nonzero energy flux out of the breather core.
The question we want to pose is whether this energy flux
can be sustained if the perturbation we choose is local in space.
The perturbation will have some overlap with the extended ones.
So there will be some radiation, but at the same time the initial
localized perturbation will simply disperse away from the breather
lowering its amplitude. The question then is whether these two
counteracting processes balance each other or not.
The question is of relevance also in connection with recently
discussed radiation mechanisms of strongly excited quantum breathers
\cite{Hizhnyakov}.
We will provide with answers for both cases.

\section{The case of classical breathers}

For the sake of simplicity we consider
first a one-dimensional lattice with the equations of motion
\begin{equation}
\ddot{x}_l+ V'(x_l) + W'(x_l - x_{l-1}) - W'(x_{l+1}-x_l) =0\;,
\label{1-1}
\end{equation}
which corresponds to the Hamiltonian
\begin{equation}
H=\sum_{l} \left[ \frac{1}{2} \dot{x}^2_l + V(X_l) + W(x_l - x_{l-1})
\right] \;.
\label{1-2}
\end{equation}
The index $l$ denotes the lattice site, and can run over a finite
or infinite lattice. Extensions to higher lattice dimensions
are straightforward. A local minimum energy state $x_l=\dot{x}_l=0$
is provided by $V(0)=W(0)=V'(0)=W'(0)=0$ and $V''(0),W''(0) > 0$.
Small amplitude excitations can be obtained by linearizing (\ref{1-1})
with the ansatz
\begin{equation}
x_l(t) \sim {\rm e}^{i(\omega_qt-ql)}
\label{1-2b}
\end{equation}
which results in the dispersion relation for plane waves
\begin{equation}
\omega_q^2 = V''(0) + 4W''(0) \sin^2 \frac{q}{2} \;.
\label{1-3}
\end{equation}
For varieties of anharmonic potentials $V,W$ it is well known
\cite{DB_rev}
that the equations (\ref{1-1}) allow for families of discrete breather
solutions of the type
\begin{equation}
\bar{x}_l(t)=\bar{x}_l(t+T_b)\;,\; x_{|l| \rightarrow \infty} \rightarrow 0\;.
\label{1-4}
\end{equation}
Here the breather frequency $\Omega_b=2\pi/T_b$ is a tunable
parameter which satisfies the nonresonance condition
\begin{equation}
m\Omega_b \neq \omega_q\;.
\label{1-5}
\end{equation}
In addition given a breather family we can generate new
families by discrete translations $l \rightarrow l + l_0$.

In the next step we consider small perturbations $\epsilon_l$ around
a given breather solution for a given value of its frequency
$\Omega_b$. We insert the ansatz $x_l(t)=\bar{x}_l(t) + \epsilon_l(t)$
into the equations of motion (\ref{1-1}) and linearize it with respect
to the perturbation:
\begin{equation}
\ddot{\epsilon}_l + V''(\bar{x}_l(t))\epsilon_l
+ W''(\bar{x}_l(t) - \bar{x}_{l-1}(t))(\epsilon_l-\epsilon_{l-1}) -
W''(\bar{x}_{l+1}(t)-\bar{x}_l(t))(\epsilon_{l+1}-\epsilon_l) =0
\;.
\label{1-6}
\end{equation}
Integration of these Floquet equations over one breather period $T_b$
maps the phase space $(\epsilon_l,\dot{\epsilon}_l)$ onto itself
and is equivalently described by a Floquet matrix $F$.
This matrix is symplectic and can be obtained e.g. numerically.
Its eigenvalues $\lambda$ and
eigenvectors $\vec{y}_{\lambda}$ describe the linear stability
properties of the DB and the scattering of plane waves by the DB as well.
If $|\lambda|=1$ the corresponding eigenstate $y_{\lambda}$ is marginally
stable - it neither grows nor decays in time.
Eigenvectors may be localized or extended. Since the breather
is localized, the eigenvalue spectrum of the localized eigenstates
is discrete, while the eigenvalue spectrum of delocalized eigenstates
is continuous for an infinite lattice. Indeed, in that case
delocalized eigenvectors far from the breather asymptotically
take the form (\ref{1-2b}). Consequently their eigenvalues are
given by
\begin{equation}
\lambda_q = {\rm e}^{i \omega_q T_b}\;.
\label{1-7}
\end{equation}
Note that this is true only for an infinite lattice. In the following
we will remind the reader about finite size corrections to this picture.

Let us now consider a lattice with a particular breather solution
such that
\begin{equation}
\omega_q \pm \omega_{q'} = m \Omega_b\;,\; q \neq q'\;,
\label{1-8}
\end{equation}
so that $\lambda_q=\lambda_{q'}$.
This twofold degeneracy
is the origin of inelastic multichannel scattering \cite{Cretegny}
performed
at these wavenumbers $q$ and $q'$.
Note, that such a situation corresponds to the parametric
resonance in the system (\ref{1-6}), where the breather acts as a parametric
driver (exponentially localized in space). Thus an important question
arises: can one pump energy into the system in the regime of a
parametric resonance, provided that parametric driving is local in space?
The answer is no, as long as we deal with an infinite system size.
Here however we note, that
for a finite lattice the outcome will be close to the above
statements, but not identical. Indeed, as shown by Mar\'in and Aubry
\cite{Marin_Aubry},
the degeneracy of the corresponding Floquet eigenvalues is lifted in a
finite lattice, leading to their departure from the unit circle,
so that $|\lambda| \neq 1$. However these departures are the smaller
the larger the lattice size is. Mar\'in and Aubry have shown using
a band analysis that $||\lambda|-1| \sim 1/N$, where
$N$ is the total number of lattices sites.
This result is obtained first by noting that the twofold degeneracy
is lifted due to the time-periodic localized DB perturbation
and can be accounted for by estimating the corresponding matrix
element. Because both initially degenerate eigenstates are delocalized,
whereas the DB is localized, the matrix element will be of the order of
$1/N$. Perturbation theory of two degenerate states tells immediately
that the degeneracy will be lifted to an amount of $1/N$ as well.

It is instructive
to revisit here the consequencies. Assume we have a
large but finite lattice and such an extended slightly unstable
eigenstate. Taking the perturbation along this eigenstate, and
integrating over one period of the breather $T_b$, this
perturbation will grow everywhere in the system, though not very strongly.
How can that happen, if the breather itself is localized, say
exponentially? The influx of energy is provided
by the breather, and is confined to a finite part of
the lattice. The only possibility is that there is a nonzero
energy flux in the outer regions of the lattice due to some
slightly inhomogeneous profile of the Floquet eigenstate.
Then the breather simply feeds energy in (or out) of a
confined small part of the lattice, and this energy then travels
along the rest of the lattice. If so, the legitimate question
arises whether this can be used as a possible source of
radiation of waves by a breather, if a corresponding generic
localized perturbation of the breather is excited.

To answer this question let us assume that all the elements of
a given (extended) Floquet eigenvector are of the same order of magnitude
$\vec{y}_{\lambda} \sim A$.
Then taking an initial perturbation being equal to such an eigenvector
the stored energy is given by $E_{\lambda} \sim N A^2$.
If we choose a growing unstable eigenvector, after one period $T_b$
the amplitude of the perturbation will become $(1+1/N)A$.
Consequently the energy after that time increases to
$N(1+1/N)^2 A^2$. Thus the energy grows during that time
by an amount of $\Delta E \sim 2A^2$ for large $N$.
Assume now that we make a local perturbation of the breather
given by some vector $\vec{B}$ with its elements again being at most
of order $A$. Then it can be represented as a superposition of
the Floquet eigenvectors:
\begin{equation}
\vec{B} \sim \frac{1}{N} \sum_{\lambda} \vec{y}_{\lambda}
\;.\label{1-9}
\end{equation}
Consequently the growth of energy in each unstable eigenstate
during one period $T_b$ of the breather amounts to
$\Delta E_{\lambda} \sim A^2/N^2$. The total number of unstable
eigenstates will be always less than $N$, so that the total
growth of energy in the direction of all unstable eigenstates
is limited by $A^2/N$. Recalling that $A$ is fixed,
this growth rate will scale to zero for an infinite lattice.
Consequently we predict that the above discussed finite lattice instabilities
do not result in an energy radiation of the breather when
perturbed locally. In other words, a local perturbation will
start to disperse away from the breather, lowering its
amplitude in the breather core. At the same time it has some
overlap with unstable eigenstates which would result in
a growth of the same amplitude. However the dispersion acts
more efficiently, and for large times the amplitude
of the perturbation in the breather core will ultimately decay down to zero.

The above argumentation was provided for a one-dimensional system.
It is straightforward to generalize it to higher lattice dimensions $d$
as well. Assuming now that $N$ represents the linear size of the
system, the stored energy in an unstable eigenvector becomes
$E_{\lambda} \sim N^d A^2$. After one period $T_b$ the energy growth
will be given by $\Delta E \sim 2 N^{d-1}A^2$. For a local perturbation
the growth of energy in each unstable eigenstate per $T_b$ is
then given by $\Delta E_{\lambda} \sim 2 N^{-d-1}A^2$. Since there are at most
$N^d$ unstable eigenvectors, we obtain again an upper limit for
the total energy growth $A^2/N$, independent of the dimension of the lattice.

\begin{figure}
\includegraphics[angle=270, width=0.47\textwidth]{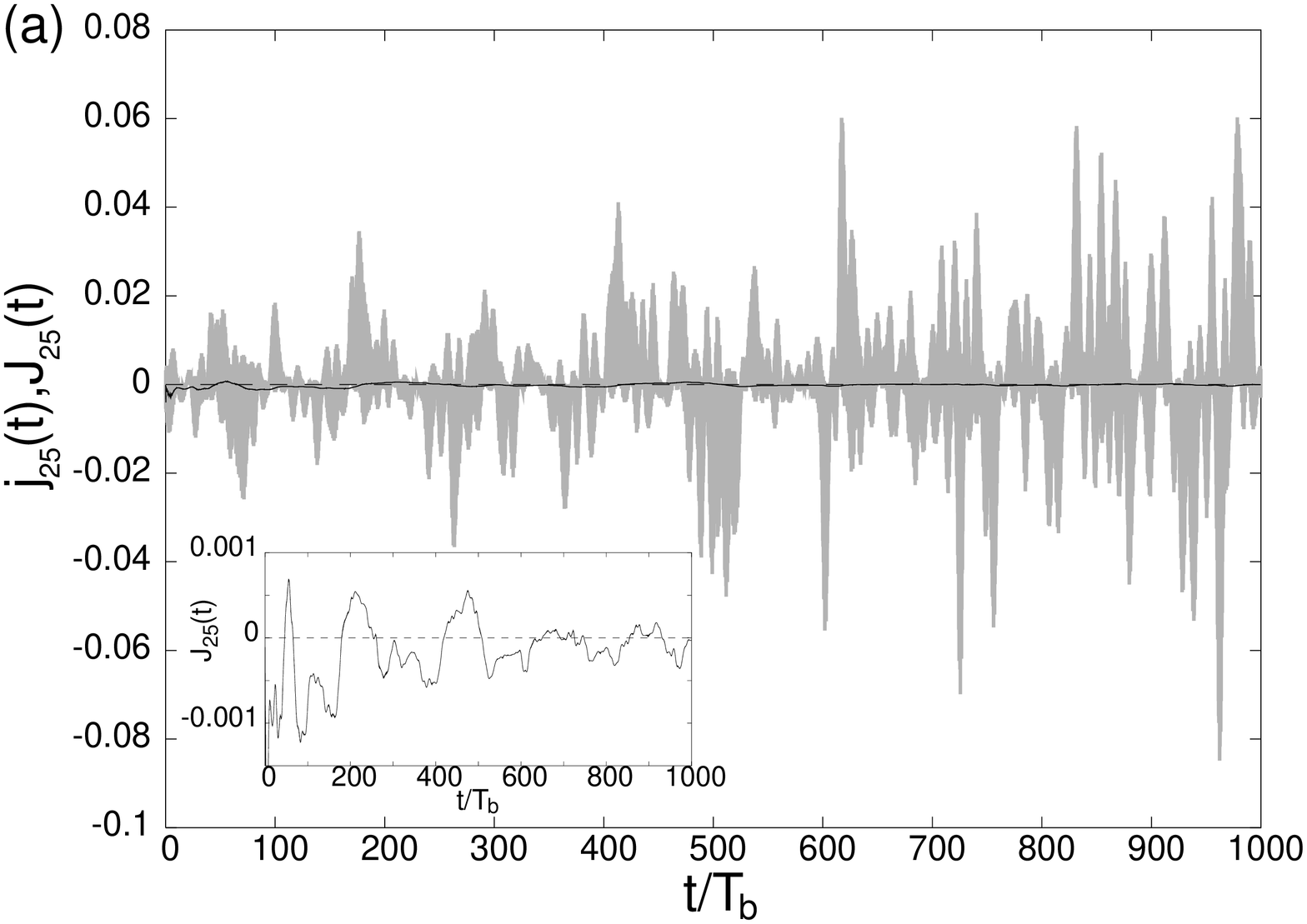}
\includegraphics[angle=270, width=0.47\textwidth]{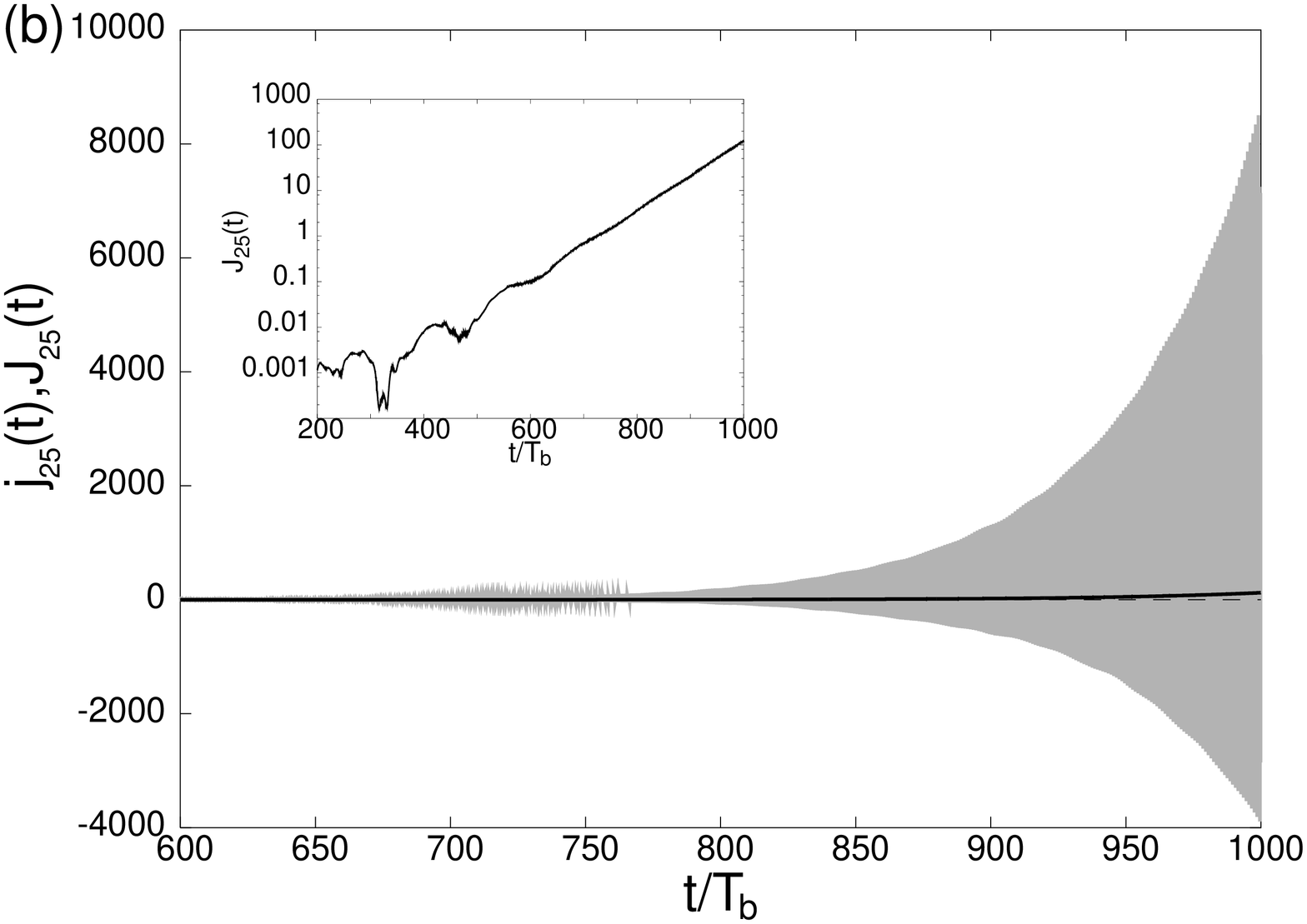}
\caption{Energy flux $j_l(t)$ (solid gray lines) and integral
energy flux $J_l(t)$
(solid black lines, see also insets) in breather dynamics with
small perturbation for
the cases: (a) $C=0.092$, perturbation along stable eigenstate;
(b) $C=0.093$,
perturbation along unstable eigenstate. The system size is $N=40$
(breather is centered
at $20$th site), energy flux is monitored at the site $l=25$.
}
\label{fig1}
\end{figure}

In order to verify this prediction, we performed high precision
numerical simulations. We consider a system with $V(x)=x^2/2-x^3/3$
and $W(x)=0.5C x^2$. First we consider a lattice of size $N=40$.
We identify a breather with frequency $\Omega_b=0.75$
for which the abovementioned finite lattice instabilities take place in
the coupling constant region $C\gtrapprox 0.093$.
We compute the eigenvalues and eigenvectors of the corresponding Floquet
matrix for two different values of the coupling constant: $C=0.092$
(breather is linearly
stable) and $C=0.093$ (breather has finite size instabilities).
We then perturb the breather in the direction of a stable (at $C=0.092$) and
a slightly unstable (at $C=0.093$)
extended eigenstates and monitor the energy flux
at some distance from the breather

\begin{equation}
j_l(t) = \dot{\epsilon}_l ( \epsilon_{l-1} - \epsilon_{l+1})
 \label{1-10}
\end{equation}
as well as its integral over the observed time of simulation $\tau$
\begin{equation}
J_l(\tau) = \frac{1}{\tau} \int_0^{\tau} j_l(t) dt
\;.
\label{1-11}
\end{equation}
We nicely reproduce the expected two different scenaria: oscillations
of energy flux with
zero average of its integral value in the case of a stable breather; and
slow but still exponential
growth of the  energy flux out of
the breather region in the case of an unstable breather, see Fig.~\ref{fig1}.
%
%
\begin{figure}
\includegraphics[angle=270, width=0.4\textwidth]{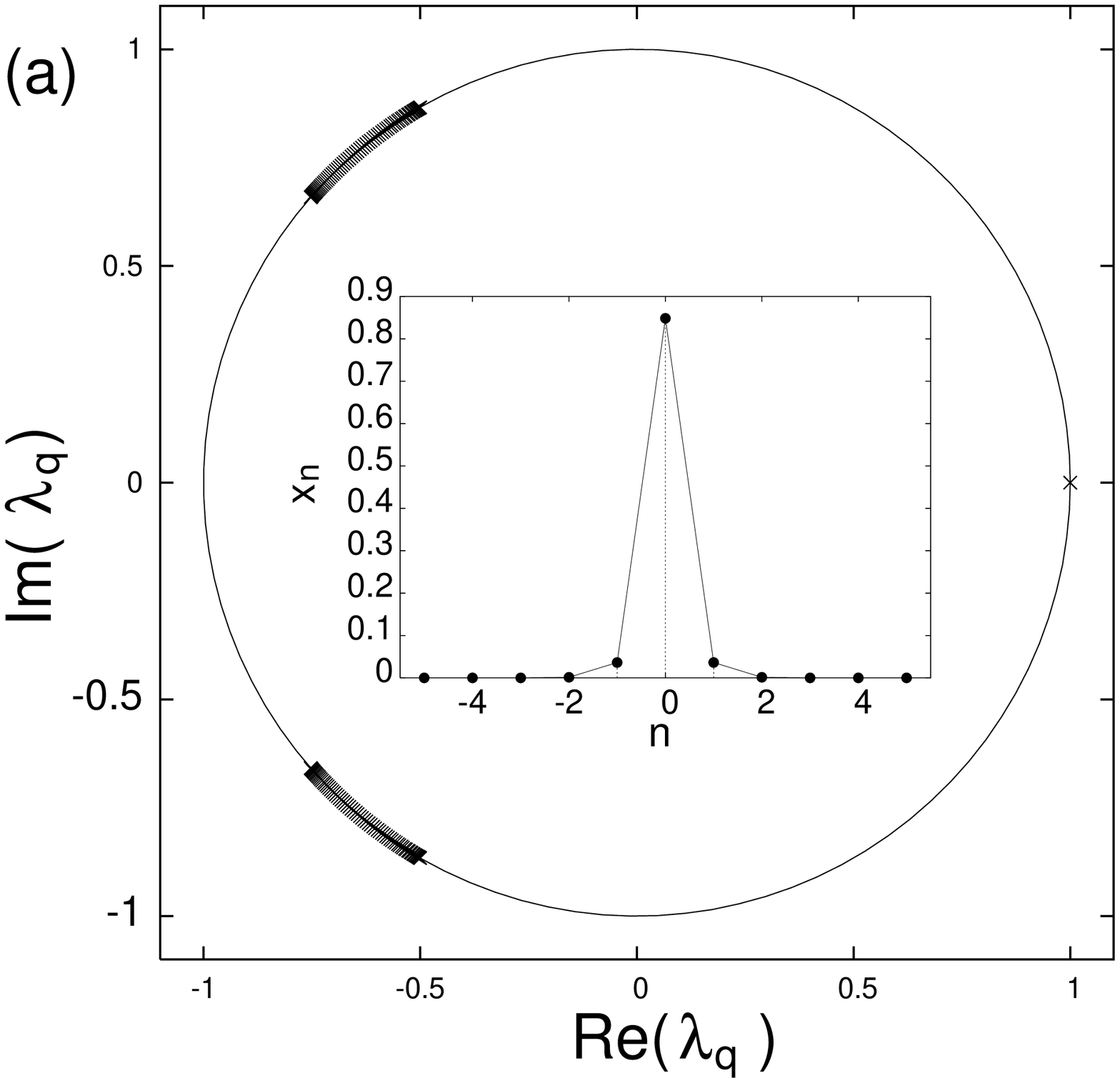}
\includegraphics[angle=270, width=0.4\textwidth]{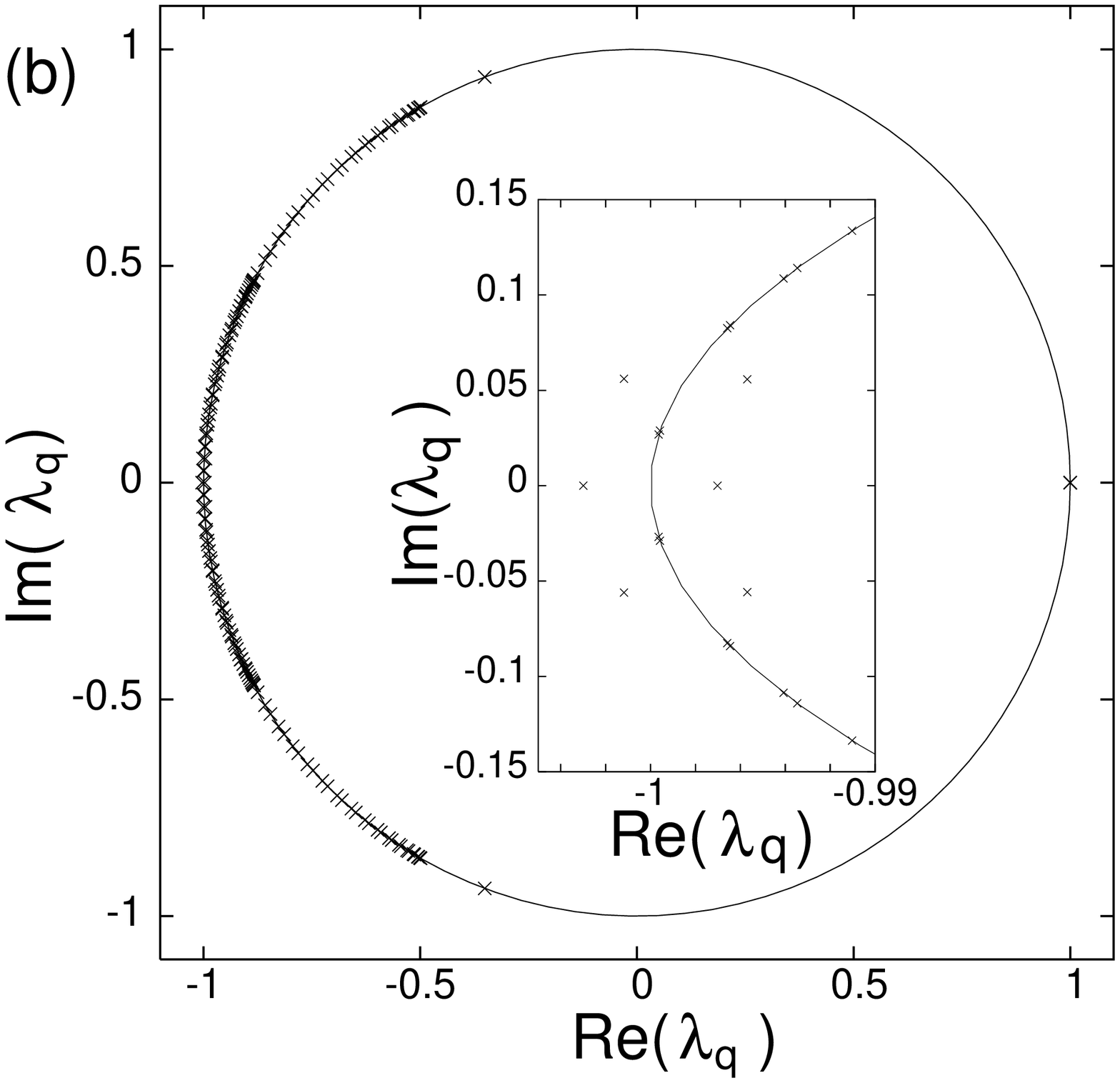}
\caption{Eigenvalues $\lambda_q$ of the Floquet matrix $F$ for the cases:
(a) $C=0.02$, $\Omega_b=0.75$, inset shows the breather profile (central part);
(b) $C=0.1$, $\Omega_b=0.75$, inset zooms the region of instabilities.
The system size in both cases is 80 sites.
}
\label{figBR}
\end{figure}

%
\begin{figure}
\includegraphics[angle=270, width=0.47\textwidth]{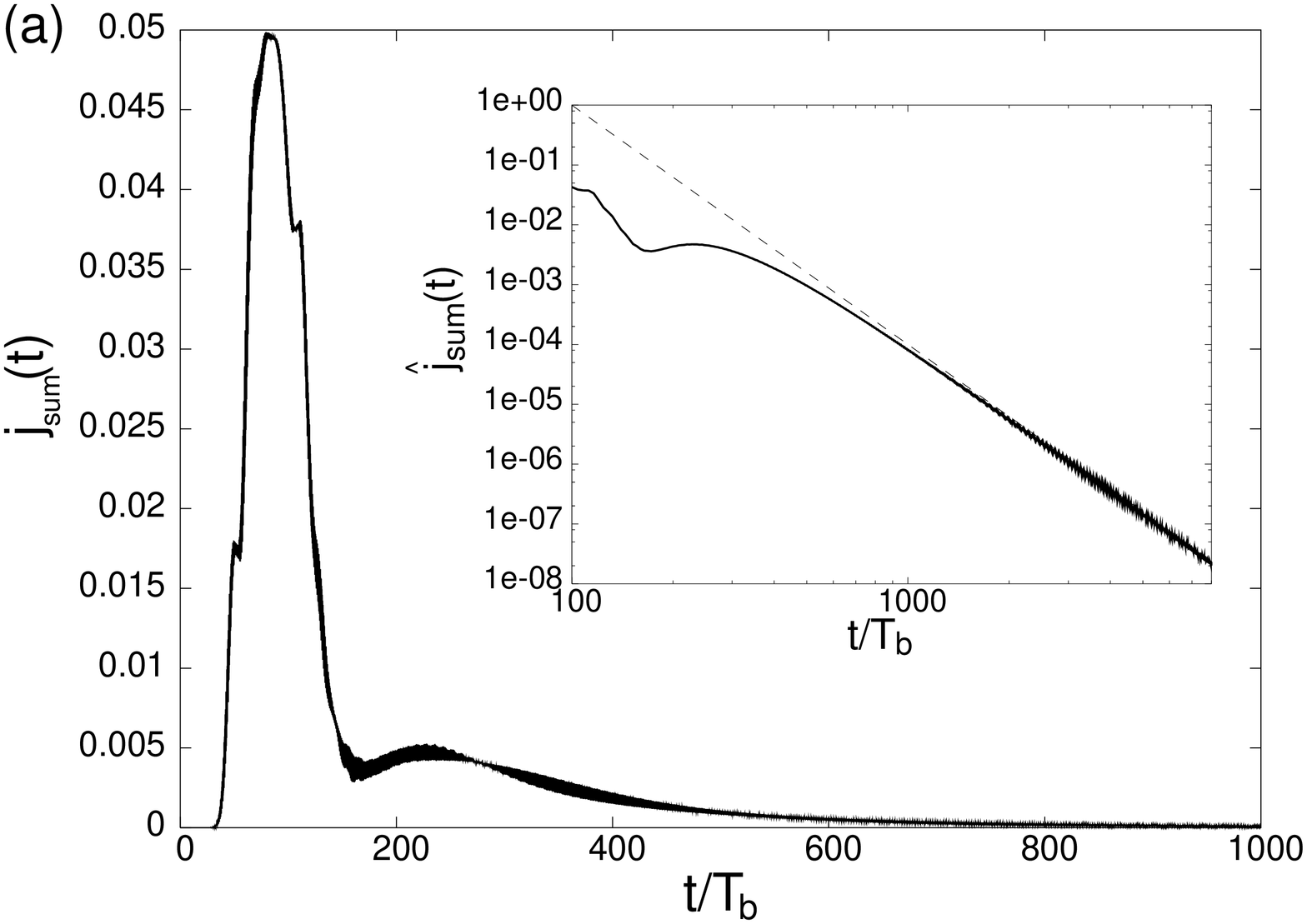}
\includegraphics[angle=270, width=0.47\textwidth]{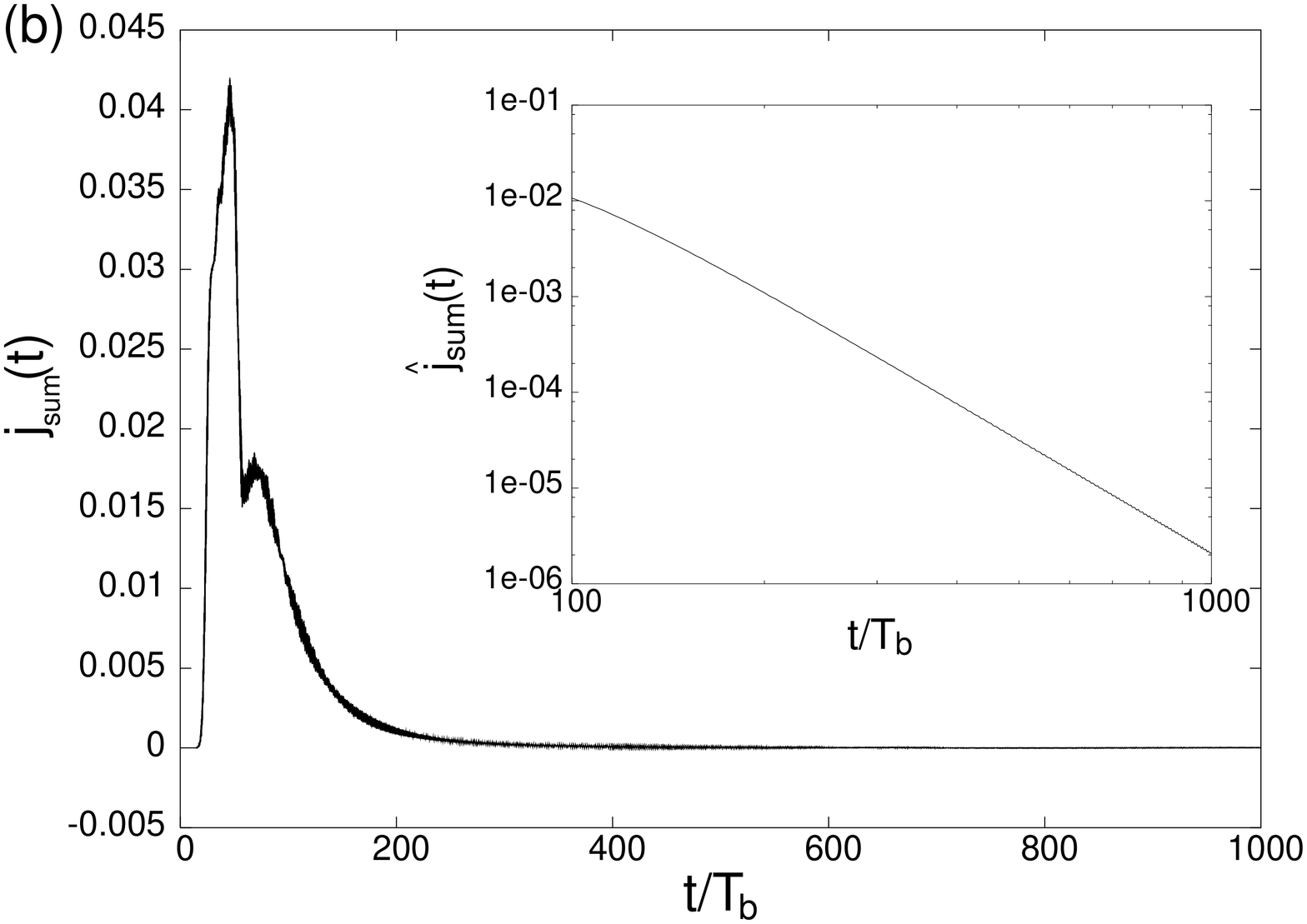}
\caption{Energy flux $j_{sum}(t)$ for a local initial perturbation for
the cases: (a) $C=0.02$; (b) $C=0.1$. The system size is $N=4000$ (breather is centered
at $2000$th site), energy flux is monitored at sites $l\in[2020,2040]$,
perturbation is made on top of the 10 adjacent sites in the breather core
in a random way. Insets show the averaged over 4 breather periods integral energy flux
$\hat{j}_{sum}(t)$ (see the main body text for details). Dashed line in inset in (a) shows the asymptote
$\hat{j}_{as}(t)=1e$+8$/(t/T_b)^4$.
}
\label{fig2}
\end{figure}
Next we take a large lattice with $N=4000$ and identify breather states
with $\Omega_b=0.75$ for
two different values of the coupling constant:
$C=0.02$ (the breather is linearly stable, see Fig.~\ref{figBR}(a))
and $C=0.1$
(there exists a considerable number of
unstable extended eigenstates, see Fig.~\ref{figBR}(b)). We excite a local perturbation
on top of the breather core in a random way on 10 adjacent sites
with amplitude being of order one at each site. We then monitor
the time dependence of the energy flux averaged over 20 lattice sites
$j_{sum}(t)=\sum_{n}^{n+20}j_l(t)$
outside the breather core.
In both cases we observe
a decay of the measured radiation
with growing time, see Fig.~\ref{fig2}. For large times it follows
the power law $j\sim t^{-4}$ (see insets in Fig.~\ref{fig2}, to smoothen
the oscillations and make the picture more clear in insets
we plot the
flux value averaged over 4 breather periods:
$\hat{j}_{sum}(t)=1/(4T_b)\int_{t-4T_b}^{t}j_{sum}(\tau)d\tau$),
which is explained in the Appendix.
This clearly demonstrates that the finite lattice
instabilities are not capable of sustaining a nonzero energy
radiation of the breather induced by a local perturbation.

\section{The quantum case}

The general problem of quantizing (\ref{1-2}) and understanding
the correspondence between quantum eigenstates and classical
discrete breathers is an issue of current research.
Since the
underlying models are nonintegrable in general, no analytic solutions
can be obtained. Approximations have to be complemented by
numerical studies. However this is in general very hard to achieve,
since one has to deal with the diagonalization of huge matrices.
Indeed, even when studying one single oscillator, strictly speaking
one has to consider an infinite dimensional Hilbert space. So
already at the level of a single oscillator justified cutoffs in
its Hilbert space have to be introduced. Taking into account
many coupled oscillators leads to the necessity to reduce the
number of states per oscillator. Reliable results in the high energy
domain of the quantum problem have been obtained so far only for
small systems of two or three coupled oscillators
\cite{Quant_Dimer, Quant_Trimer, Quant_Tun, Quant_Tun_Roto}                            .
These studies
together with computations of larger systems at lower energies
have confirmed that quantum breather states are many phonon bound states
\cite{Eilbeck}.
They correspond to some extent to classical breather excitations being
able to tunnel along the lattice
\cite{Quant_Tun2}.
The tunneling rate is expected
(and confirmed for small systems) to be exponentially small for
large numbers of participating phonons.
Then it is legitimate to assume that a localized excitation of
the lattice with an energy corresponding to a large number of phonons
will evolve for exponentially long times according to its classical
DB analogue. During that time however the evolution of small amplitude
perturbations around the breather can be considered in its
full quantum version. In order to proceed we observe that (\ref{1-6})
corresponds to a time-dependent Hamiltonian
\begin{equation}
H_{\epsilon}= \sum_l \left[ \frac{1}{2}\pi_l^2 + \frac{1}{2}
V''(\bar{x}_l(t))\epsilon_l^2 + \frac{1}{2}
W''(\bar{x}_l(t) - \bar{x}_{l-1}(t))(\epsilon_l-\epsilon_{l-1})^2 \right]
\;.
\label{2-1}
\end{equation}
Here $\pi_l$ is the canonically conjugated momentum to $\epsilon_l$.
Now we may consider the corresponding quantum Hamiltonian operator
\begin{equation}
\hat{H}_{\epsilon}= \sum_l \left[ \frac{1}{2}\hat{\pi}_l^2 + \frac{1}{2}
V''(\bar{x}_l(t))\hat{\epsilon}_l^2 + \frac{1}{2}
W''(\bar{x}_l(t) - \bar{x}_{l-1}(t))(\hat{\epsilon}_l-
\hat{\epsilon}_{l-1})^2 \right]
\;.
\label{2-2}
\end{equation}
The operators satisfy the standard commutation relation 
$\left[ \hat{\epsilon}_l,\hat{\pi}_m\right] = i \delta_{l,m}$.

The fact that the Hamiltonian (\ref{2-2}) is a quadratic form of
operators $\hat{\epsilon}_l$ and $\hat{\pi}_l$ is crucial: we may appeal
to the Ehrenfest theorem \cite{Ehrenfest} and conclude that the dynamics
of the quantum system (\ref{2-2}) is in full agreement with the dynamics
of the corresponding classical system (\ref{2-1}) and therefore no
additional quantum effects which may lead to breather radiation should appear.
To show this we switch to time-dependent Heisenberg operators
$\hat{\epsilon}_l^{H}(t), \hat{\pi}_l^{H}(t)$ using standard
relations:
\begin{equation}
\label{Heisenberg}
\hat{A}^{H}(t)
 =
(\tilde T e^{i\int_0^t d\tau
\hat{H}_\epsilon})
\hat{A}
(T e^{ - i\int_0^t d\tau
\hat{H}_\epsilon}),
\end{equation}
where $\hat{A}^{H}(t)$ is the time-dependent Heisenberg operator corresponding to a
time-independent operator $\hat{A}$, $T$ and $\tilde T$ are time ordering and anti-time ordering operators, respectively. The equation of motion for a Heisenberg operator
$\hat{A}^{H}(t)$ reads:
\begin{equation}
\label{Heis_eq}
-i \frac{\partial \hat A^H(t)}{\partial t} =
[\hat{H}_\epsilon(\hat\pi,\hat\varepsilon,t), \hat A^H(t)].
\end{equation}

After substitution of the Hamiltonian (\ref{2-2}) into (\ref{Heis_eq}) it follows that
operators $\hat{\epsilon}^H_l(t)$ satisfy the discussed above classical equations
(\ref{1-6}), in which
coordinates $\epsilon_l$ are substituted by the corresponding
operators $\hat{\epsilon}^H_l$. Since these equations are linear we may
average them with the time-independent wave function $\psi_0$, corresponding
to the initial state of the system, and get an
equation for the expectation value of the coordinate operator
\begin{equation}
\label{exp_val}
\tilde\varepsilon_l(t) = \langle \psi_0|\hat\epsilon_l^H(t)|\psi_0
\rangle
\end{equation}
which is identical to (\ref{1-6}) with expectation value
$\tilde\varepsilon$ standing instead of the classical variable
$\epsilon$. Therefore all the conclusions made in the previous chapter
as for time evolution of the classical coordinate
$\epsilon_l$ hold for the expectation value $\tilde\varepsilon_l$
of the quantum operator $\hat\varepsilon_l$ as well.

Alternatively, the question of possible radiation of phonons in the quantum case
may be studied by considering
a quantum Floquet problem. Let us split the Hamiltonian (\ref{2-2}) into a time-averaged
and an ac part:
\begin{equation}
\hat{H}(t) = \hat{H}_{dc} + \hat{H}_{ac}(t) \;.
\label{2-3}
\end{equation}
The ac part $\hat{H}_{ac}$ is local in space because of the locality of
the discrete breather which is the origin of the ac drive.
We can consider the full orthonormal basis of eigenfunctions $\phi_{\nu}$
of the dc part
\begin{equation}
\hat{H}_{dc} \phi_{\nu} = \epsilon_{\nu} \phi_{\nu}
\label{2-4}
\end{equation}
and expand the full wave function $\Psi$ which satisfies
the time-dependent Schr\"odinger equation
\begin{equation}
i \dot{\Psi} = \hat{H}(t) \Psi
\label{2-5}
\end{equation}
in that basis:
\begin{equation}
\Psi (t) = \sum_{\nu} C_{\nu}(t) \phi_{\nu}
\;.
\label{2-6}
\end{equation}
This will lead to a set of coupled first order differential
equations for the coefficients $C_{\nu}$:
\begin{equation}
i\dot{C}_{\nu} = \epsilon_{\nu} C_{\nu} + \sum_{\mu} h_{\nu \mu}(t) C_{\mu}
\label{2-7}
\end{equation}
where the matrix elements
\begin{equation}
h_{\nu \mu}(t) \langle \phi_{\nu} \hat{H}_{ac}(t) \phi_{\mu} \rangle
\label{2-8}
\end{equation}
have been introduced and $\langle ... \rangle$ denotes the scalar product
in the space of $\phi$.
We also note that because $\hat{H}$ is Hermitian (in fact real symmetric)
\begin{equation}
\frac{d}{dt}|\Psi(t)|^2 = \frac{d}{dt} \sum_{\nu} |C_{\nu}(t)|^2 =0\;.
\label{2-9}
\end{equation}

In order to answer the question of radiation, let us start with noticing
that (\ref{2-7}) constitute a Floquet problem similar to the classical
case. The difference is that the rank of the Floquet matrix is formally
speaking infinite even for a finite lattice, since the Hilbert
space dimension of a single site oscillator is infinite.
The extended states in (\ref{2-4}) can be characterized by the amount
of excited one-phonon energies and classified accordingly as many-phonon
excitations with a given number of phonons participating. This
constitutes the main difference to the classical Floquet problem
(\ref{1-6}), where only pairs of one-phonon excitations appear
(because in the classical case we compute frequencies instead
of energies, and time reversal symmetry provides with two possible
signs of the phonon frequency). For the quantum case we have an infinite
number of one-, two-, three-phonon excitations etc.
To estimate the magnitude of the departure of quantum Floquet eigenvalues
from the unit circle, we need again, as in the classical case, to
first estimate the matrix elements. Since the DB solution has a main frequency
contribution and higher harmonics with amplitudes exponentially decaying
with increasing order, the main contribution will originate from
the main frequency component of the DB, which couples the space
of many-phonon states locally, e.g. the
ground state with the two-phonon states,
the two-phonon states with the four-phonon
states, the four-phonon states with the six-phonon states etc.
In other words, we have to consider an infinite matrix with
degenerate diagonal elements and nearest neighbour interaction
elements of the order of $1/N$ in analogy to the classical case.
The eigenvalue spectrum of such a matrix will spread around
the value of the diagonal elements to the same order $1/N$.
Thus we conclude that the quantum Floquet eigenvalues will
depart from the unit circle not farther than $1/N$ exactly as in
the classical case.

What remains then is to repeat the final argument applied in the classical
case.
This can be done in full analogy to the previous chapter,
noting that the observable of the flux $j_l(t)$ will
be defined through the corresponding operator $\hat{j}$
and the product $j(t) = \langle \Psi|\hat{j} | \Psi \rangle$.
This product will be a quadratic form of the time-dependent
coefficients $C_{\nu}(t)$, which completes the above analogy.
The conclusion is thus that despite the fact that the
quantum Floquet problem involves an infinite number of bands,
and despite the fact that the norm is conserved in the quantum case,
a local perturbation around the breather will not lead to
a persistent radiation of phonons. The argument that radiation must take
place because the ground state of the unperturbed system (without DB)
is not anymore the ground state of the system with a DB must then be misleading.
In fact the computation of the exact quantum Floquet eigenvalue problem
will show that there is always a locally deformed ground state (as well
as the excited states). That deformation is clearly not the cause of
radiation. The only possible cause - an instability of extended Floquet
states - has been excluded by the
above reasoning.

\section{Conclusions}

In this work we excluded a particular mechanism of radiation of
perturbed breathers driven via weak finite size instabilities
of extended states. While the statement is rigorous when
treating the whole system classically, we arrive at a similar result
also when treating the fluctuations quantum mechanically, leaving the
breather solution to be a given classical one.
One way to obtain nonzero radiation is to include
higher order terms of the perturbation $\epsilon$ which together
with possible localized Floquet eigenvectors of the linearized
phase space flow will provide with a constant radiation rate
of the breather into the plane wave continuum, both for a classical
as well as a quantum treatment of the fluctuations.
Another path is to quantize the breather itself in the quantum
case. Then we can expect breather tunneling along the lattice,
which will provide with some diffusion of the full breather
energy out of the originally excited lattice part.
Concluding we may say that discrete breathers are surprisingly
robust objects. They can radiate energy into the continuum of
a large lattice only via higher orders of perturbations around them.

\vspace{1cm}

\appendix{\bf APPENDIX}

An estimation of a wave packet dynamics, resulted from a local perturbation on
a lattice, can be made essentially in a similar
way as it was done for continuous systems \cite{Kovalev}.
Let us start with an instructive case of a lattice without
any breather. We excite a local initial perturbation, say
$\epsilon_l(0)=\delta_{l,0}$. Its representation in the reciprocal
lattice space with wave number $q$ is given by a $q$-independent constant
$\epsilon_q=const$.
Consequently the evolution of the perturbation after some time $t$
will be given by
\begin{equation}
\epsilon_l(t) \sim  \int_{0}^{\pi} \epsilon_q {\rm e}^{i(ql-\omega_q t)} dq\;
\label{a1}
\end{equation}
with $\omega_q$ given by the plane waves dispersion relation (\ref{1-3}).
Rewriting it as
\begin{equation}
\int_{0}^{\pi} {\rm e}^{i(ql-\omega_q t)} dq = \int_{0}^{\pi} {\rm e}^{iF(q)t}dq
\label{a2}
\end{equation}
with $F(q)=q\frac{l}{t}-\omega_q$ we can estimate the integral by
noting that for large values of $t$ only $q$-values contribute
for which $\partial F/\partial q = \frac{l}{t} - v_q$ is small. Here $v_q$ is the group velocity
at wave number $q$. For large $t$ only waves with small group velocities
contribute, i.e. with wave numbers $q$ close to the edges of the first Brilluen
zone $q=\hat{q}=0,\pi$. It is straightforward to show, that all extremum points $q=q^*$
of the function $F(q)$
give contribution to $\epsilon_l(t)$ of the same order in small parameter $1/t$ and
do not cancel each other,
hence it is enough to consider only a single extremum point.
Expanding $F$ around its extremum $q=q^*$, the integral can be estimated
to be
\begin{equation}
\int_{0}^{\pi} {\rm e}^{iF(q)t}dq \sim
{\rm e}^{iF(q^*)t}\int
{\rm e}^{i0.5F^{''}(q^*)\left(q-q^*\right)^2t}dq
\sim
\frac{1}{\sqrt{t}}{\rm e}^{i(\xi
\frac{l^2}{t} +\hat{q} l - \omega_{q=q^*}t)}\;,
\label{a3}
\end{equation}
where $\xi=\left(dv_q/dq\right)^{-1}|_{q=\hat{q}}$.
Using the definition (\ref{1-10}) we then obtain
\begin{eqnarray}
\dot{\epsilon}_l(t)&\sim& t^{-0.5}\\
\epsilon_{l+1}(t)-\epsilon_{l-1}(t) &\sim& t^{-0.5} \sin(\hat{q}+2\xi l/t)
\sim t^{-1.5}\\
j_l(t) &\sim& t^{-2}\;.
\label{a4}
\end{eqnarray}
We tested this prediction numerically and found complete agreement.

The observed $1/t^4$ dependence
for a perturbation on top of the breather can be now explained
by noting that the breather represents a local violation of
the translational invariance.
In such a case the abovementioned reciprocal lattice
representation $\epsilon_q$ of a local perturbation becomes
$q$-dependent because plane waves are not the true eigenstates
of the system anymore. It is easy to show that already for a
single site time-averaged breather contribution the amplitude
of the extended eigenstates at the breather site is proportional to
$\sin q$. This implies that $\epsilon_q \sim \sin q$.
Following then again the above reasoning, we obtain
\begin{eqnarray}
\epsilon_l(t) &\sim& \sin(q^*){\rm e}^{iF(q^*)t}\int
{\rm e}^{i0.5F^{''}(q^*)\left(q-q^*\right)^2t}dq\\
\nonumber
&\sim& t^{-1.5}{\rm e}^{i(\xi
\frac{l^2}{t} +\hat{q} l - \omega_{q=q^*}t)},\\
j_l(t) &\sim& t^{-4}
\end{eqnarray}
confirming the numerical results in Fig.\ref{fig2}.

\end{document}